\documentclass[aps,prd,twocolumn,nofootinbib,
superscriptaddress,floatfix,preprintnumbers]{revtex4-2}

\usepackage{amsmath,amssymb,amsfonts,bm}
\usepackage{graphicx}
\usepackage{dcolumn}
\usepackage[dvipsnames]{xcolor}
\usepackage{slashed}
\usepackage[normalem]{ulem}
\usepackage{tikz}
\usepackage[colorlinks=true, linkcolor=blue, urlcolor=Blue, citecolor=Mulberry]{hyperref}
\usepackage{booktabs}
\usepackage{orcidlink}
\allowdisplaybreaks
\newcommand{\beq}{\begin{equation}\begin{aligned}}
\newcommand{\eeq}{\end{aligned}\end{equation}}

\begin{document}

\title{Does the High Energy LUX-ZEPLIN Event Suggest Hyperfine Atomic Dark Matter?}

\author{Prolay Chanda,\orcidlink{0000-0002-3940-6062}}
\email{chandakr@ualberta.ca}
\affiliation{Department of Physics, University of Alberta, CCIS 4-181, Edmonton, Alberta T6G 2E1, Canada}
\affiliation{Arthur B. McDonald Canadian Astroparticle Physics Research Institute, 64 Bader Lane, Queen's University, Kingston, ON Canada, K7L 3N6}

\author{Bhaskar Das\,\orcidlink{0009-0006-9881-5401}}
\email{bdas4@uic.edu}
\affiliation{Department of Physics, University of Illinois Chicago, Chicago, IL 60607, USA}

\author{Sagnik Mukherjee\,\orcidlink{0009-0000-9375-225X}}
\email{smukhe34@uic.edu}
\affiliation{Department of Physics, University of Illinois Chicago, Chicago, IL 60607, USA}

\begin{abstract}
 The LUX-ZEPLIN (LZ) Collaboration has reported a nuclear-recoil candidate at $248\pm23~({\rm stat})\pm23~({\rm sys})~{\rm keV}$ in an extended high energy search window. We investigate whether this event can arise from the hyperfine excitation of hydrogen-like atomic dark matter. For a TeV-scale dark atom, a hyperfine splitting near $340~{\rm keV}$ places xenon scattering close to the observed value by naturally selecting the high-velocity tail of the Galactic halo while suppressing the leading low-energy elastic response. The resulting endothermic kinematics predict a pronounced target hierarchy: the benchmark transition is inaccessible on Ar and Ge, lies close to threshold on Xe, and remains open on heavier targets such as W, which is specific to our model parameters. Because xenon probes the extreme high-speed tail, the signal exhibits a large annual modulation and a strong dependence on the assumed halo distribution. 
We also examine how the infall of the Large Magellanic Cloud enhances the high-speed tail of the Milky Way's local dark matter distribution, extending the kinematic reach to larger hyperfine splittings.

\end{abstract}

\maketitle

\section{Introduction}

The particle nature of dark matter remains one of the most challenging problems in physics. Despite a growing number of experimental and observational searches, none has yet produced an unambiguous signal of a dark matter interaction. But very recently, the LUX-ZEPLIN (LZ) experiment, by extending its nuclear-recoil search to higher energies, has reported a captivating result - a nuclear recoil candidate with a reconstructed recoil energy
$E_R=248\pm23~({\rm stat})\pm23~({\rm sys})~{\rm keV}$ \cite{LZ:2026axp}. 

In the minimalistic realization for elastic WIMP-nucleon scattering, the probability of a recoil energy event above 200 keV is sub-percent.  Hence, various dark matter interpretations \cite{Yin:2026jnn, Fan:2026kxx, Wu:2026nhi, Freese:2026sga, Su:2026rwz, Yamashita:2026ump, Visinelli:2026kgt, Sheng:2026tqt, Sannino:2026hkc,  DiMauro:2026ldr, McCabe:2026crm, Du:2026guj, Barman:2026omh, Smirnov:2026aqk, Dent:2026bji, Baer:2026fpy, Lee:2026wof, He:2026idw, Das:2026uyy, De:2026win, Yang:2026wpb, Wang:2026ytg, Bisal:2026khf, Borah:2026zwf, Bandyopadhyay:2026gjw, Du:2026lpa, deLima:2026shq,  Ahmed:2026qjg, Okada:2026eol, Langhoff:2026ujr, Lee:2026xxh, Asadi:2026iot, Lee:2026jxl, Zhu:2026dag, Yuan:2026djt, Kumar:2026lgi, Fan:2026hzw, Lian:2026hpm, Murayama:2026apt, Kalita:2026hvb,Ahmed:2026com,Ahmed:2026kan,Jeesun:2026vzo,Kotlarski:2026pep,Lou:2026idn,Das:2026jsq,Nagata:2026pbj,Oliveira:2026jtn,Brdar:2026ukx,Mahapatra:2026glu, Palmisano:2026kuj,Pallis:2026jdc,Okada:2026fef,Dutra:2026fng, Girmohanta:2026qzo,Delepine:2026ith,Gemmell:2026yaw,Maier:2026qpr,Paul:2026okh,Baer:2026yrt,Luengas:2026tlf,Borah:2026ris,Meissner:2026ooy,Unwin:2026rdp, Kannike:2026qyl, Liang:2026coz, Alhazmi:2026efz, Elahi:2026vlm,Bose:2026szs,Chauhan:2026udz, Mahapatra:2026glu,Khan:2026osp,CentellesChulia:2026ats,Jung:2026otm,Arcadi:2026kev} which were proposed soon after the LZ result came out mostly focused on inelastic dark matter \cite{Yin:2026jnn, Fan:2026kxx, Wu:2026nhi, Freese:2026sga, Su:2026rwz, Yamashita:2026ump, Barman:2026omh, Visinelli:2026kgt, De:2026win, DiMauro:2026ldr, McCabe:2026crm, Du:2026guj, Smirnov:2026aqk, deLima:2026shq, Dent:2026bji, Baer:2026fpy, He:2026idw, Lee:2026wof, Das:2026uyy, Yang:2026wpb, Wang:2026ytg, Bisal:2026khf, Borah:2026zwf, Bandyopadhyay:2026gjw, Du:2026lpa, Ahmed:2026qjg, Okada:2026eol, Langhoff:2026ujr, Lee:2026xxh, Asadi:2026iot, Lee:2026jxl, Zhu:2026dag, Yuan:2026djt, Kumar:2026lgi, Fan:2026hzw, Lian:2026hpm, Murayama:2026apt,Oliveira:2026jtn,Nagata:2026pbj, Jeesun:2026vzo,Lou:2026idn,Das:2026jsq,Kalita:2026hvb,Ahmed:2026com,Ahmed:2026kan,Kotlarski:2026pep,Brdar:2026ukx, Sheng:2026tqt, Sannino:2026hkc, Palmisano:2026kuj,Mahapatra:2026glu,Pallis:2026jdc,Dutra:2026fng,Girmohanta:2026qzo,Delepine:2026ith,Gemmell:2026yaw,Maier:2026qpr,Paul:2026okh,Baer:2026yrt,Okada:2026fef,Luengas:2026tlf,Borah:2026ris}. However, some elastic realizations have also appeared \cite{Meissner:2026ooy,Unwin:2026rdp, Kannike:2026qyl, Alhazmi:2026efz, Elahi:2026vlm,Chauhan:2026udz,Bose:2026szs,Liang:2026coz,Khan:2026osp,CentellesChulia:2026ats,Jung:2026otm,Arcadi:2026kev}, and, a similar event was proposed earlier in Ref.~\cite{Meissner:2023jwo}, for supermassive charged gravitinos. At such recoil energies, conventional coherent spectra are increasingly suppressed by the xenon nuclear form factor, while momentum-dependent and inelastic interactions can place relatively more signal weight at high energy. Endothermic dark matter provides a particularly sharp realization: part of the incident kinetic
energy is spent on an internal excitation, so a sufficiently large gap selects the high-speed tail and reshapes the accessible recoil spectrum. Among the endothermic inelastic interpretations, considerable attention has been given to the thermal Higgsino in particular, but it seems to be excluded by IceCube neutrino flux limits from dark matter capture and annihilation in the sun \cite{Pospelov:2026ewn, DiMauroShaikh:2026solar, Bose:2026ndd, Nguyen:2026lui}, and possibly also by the prediction of dark matter induced recoil events in the high-energy sideband, which the LZ does not observe \cite{Rodd:2026tyn} (non-thermal higgsinos, however, remains a possibility \cite{Rodd:2026tyn, Langhoff:2026ujr}). 

A natural setting for inelastic dark matter is composite dark matter, whose internal excitations provide the required energy splitting. Composite models within a confining dark-QCD framework have been explored in Refs.~\cite{Sheng:2026tqt,Sannino:2026hkc}. In this work, we consider a different realization: a weakly coupled, hydrogen-like dark atom. Such an atom possesses hyperfine singlet and triplet states, with a splitting determined by the constituent masses and dark fine-structure constant that fix its binding energy and Bohr radius~\cite{Kaplan:2009de,Cline:2021itd,Cline:2014eaa, Qiu:2023bbp}. Atomic structure therefore supplies a microscopic origin for the inelastic gap rather than introducing it as an independent parameter. Whether this excitation dominates direct detection depends on the interaction with ordinary matter. We consider a $U(1)$ axial portal connecting the dark and the visible sectors, for which the leading singlet elastic matrix element vanishes, while the singlet-triplet transition is allowed. The combination of atomic structure and this selection rule thus realizes a leading endothermic signal.

For a TeV-scale atom, we calculate the hyperfine recoil spectrum and use public LZ information to determine the portal coupling required for one expected event. We also examine the target dependence and annual modulation of the signal, together with its sensitivity to the halo velocity distribution. In particular, we consider a Milky Way (MW) halo distribution that includes the influence of the Large Magellanic Cloud (LMC), denoted MW+LMC. Its extended high-speed tail can make larger hyperfine splittings accessible and alter the reconstructed recoil spectrum. Finally, we assess the cosmological consistency of our direct-detection benchmark by considering dark recombination and radiation drag for an assumed thermal history, and estimate the neutral-atom self-scattering cross section.

The remainder of this paper is organized as follows. In Sec.~\ref{sec:model}, we introduce the atomic dark matter framework and its $U(1)$ portal to the visible sector. In Sec.~\ref{sec:lz}, we reconstruct the LZ signal within this model. The impact of the local dark matter velocity distribution is discussed in Sec.~\ref{sec:halo}, with particular emphasis on the high-velocity tail induced by the MW+LMC system. In Sec.~\ref{sec:cosmology}, we examine the cosmological consistency of the benchmark parameter space.

\section{Atomic dark matter and hyperfine scattering}
\label{sec:model}

We consider two Dirac fermions, a dark proton $p_D$ and a dark electron $e_D$, carrying opposite unit charges under an unbroken $U(1)_D$. The associated massless dark photon $A_D$ mediates the Coulomb interaction that binds them into a neutral dark hydrogen-like state $H_D=(p_D e_D)$ \cite{Kaplan:2009de}.

We define the constituent mass ratio and reduced mass as
\beq
r\equiv\frac{m_{p_D}}{m_{e_D}},\qquad \mu_D\equiv\frac{m_{p_D}m_{e_D}}{m_{p_D}+m_{e_D}},
\label{eq:definitions}
\eeq
with $m_D=m_{p_D}+m_{e_D}-B_D\simeq m_{p_D}+m_{e_D}$ for $B_D\ll m_D$. The leading hydrogenic binding energy and Bohr radius are \cite{Kaplan:2009de}
\beq
B_D=\frac12\alpha_D^2\mu_D,\qquad a_D^{-1}=\alpha_D\mu_D,
\label{eq:binding}
\eeq
where $\alpha_D$ is the dark fine-structure constant, $\mu_D$ is the reduced mass of the dark proton plus the dark electron system, and $a_D$ denotes the dark Bohr radius.

For pointlike Dirac constituents with gyromagnetic factors are $g_e\simeq g_p\simeq2$, and the leading $1s$ hyperfine splitting is \cite{Cline:2021itd}
\beq
\delta_{\rm hf}=\frac23g_eg_p\alpha_D^4\frac{\mu_D^3}{m_{e_D}m_{p_D}}\simeq\frac83\alpha_D^4m_D\frac{r^2}{(1+r)^4},
\label{eq:hfsplit}
\eeq
where the last expression uses $r\equiv m_{p_D}/m_{e_D}$ and neglects corrections of relative order $B_D/m_D$ in $m_D\simeq m_{p_D}+m_{e_D}$.
Thus, once $(m_D,r,\alpha_D)$ are specified, the hyperfine gap is fixed by the same microscopic parameters that determine the atomic binding energy and size \cite{Kaplan:2009de,Cline:2021itd}.

Throughout the main text we adopt the benchmark
\beq
m_D=1~{\rm TeV},\qquad r=10,
\label{eq:massbench}
\eeq
with $\delta_{\rm hf}=340~{\rm keV}$ as the central benchmark. Using eqns.~\eqref{eq:definitions}--\eqref{eq:hfsplit} we obtain the following numerical estimates for the dark atom with $\delta_{\rm hf}=330\text{--}350~{\rm keV}$ as the hyperfine energy range, 
\beq
\alpha_D\simeq0.06524\text{--}0.06621,~~B_D\simeq0.176\text{--}0.181~{\rm GeV}.
\label{eq:atomicrange}
\eeq
At the central point $\delta_{\rm hf}=340$, neglecting the $B_D/m_D$ correction to the constituent masses, we obtain,
\beq
m_{p_D}&\simeq909.1~{\rm GeV},~~ m_{e_D}\simeq90.9~{\rm GeV},~~
\mu_D\simeq82.64~{\rm GeV},\\ \alpha_D&\simeq0.06573,~~
B_D\simeq0.1785~{\rm GeV},
\label{eq:centralatom}
\eeq
which maintains the hierarchy $\delta_{\rm hf}\ll B_D\ll m_D$. 
\subsection{Axial portal}

To couple the dark atom to ordinary matter, we adopt the axial-vector portal of Ref.~\cite{Kaplan:2009de}. The binding interaction and the visible-sector portal are distinct. An unbroken $U(1)_D$, with massless gauge boson $A_D$, binds the dark
proton $p_D$ and dark electron $e_D$ into the neutral atom $H_D$. A second, broken $U(1)_X$ contains a massive gauge boson $X_\mu$ that mediates interactions with the Standard Model. 

The $U(1)_X$ field strength kinetically mixes with the Standard Model hypercharge field strength,
\beq
{\cal L}_{\rm mix}=\epsilon_Y X_{\mu\nu}B^{\mu\nu},
\label{eq:kineticmixing}
\eeq
where $\epsilon_Y$ denotes the hypercharge-mixing parameter. After electroweak symmetry breaking, diagonalization of the gauge kinetic terms, and integrating out the $Z$ boson at direct-detection momentum transfers, the relevant interactions, to leading order in the mixing and in $m_X^2/m_Z^2$, are
\beq
\begin{aligned}
{\cal L}_{\rm portal}\supset {}&
X_\mu J_D^\mu-\epsilon_Yc_WX_\mu J_{\rm EM}^\mu
-\epsilon_Ys_W\!\left(\frac{m_X}{m_Z}\right)^2X_\mu J_Z^\mu\\
&-\frac{\epsilon_Ys_W}{m_Z^2}J_{D\mu}J_Z^\mu .
\end{aligned}
\label{eq:portalLagrangian}
\eeq
The two dark constituents $p_D$ and $e_D$ couple axially to $X_\mu$ with opposite signs:
\beq
J_D^\mu=-g_5\,\bar p_D\gamma^\mu\gamma^5p_D+g_5\,\bar e_D\gamma^\mu\gamma^5e_D ,
\label{eq:darkaxialcurrent}
\eeq
where $g_5$ is the axial $U(1)_X$ gauge coupling.
Here $s_W\equiv\sin\theta_W$ and $c_W\equiv\cos\theta_W$, where the Standard Model electromagnetic and neutral currents are defined by ${\cal L}_{\rm SM}\supset A_\mu J_{\rm EM}^\mu+Z_\mu J_Z^\mu$. The low-energy photon-mixing parameter is defined as
$\epsilon_\gamma\equiv\epsilon_Yc_W .$
For the sub-GeV mediator considered here, the propagating $X$ interaction is electromagnetic-current dominated. At $m_X=0.20~{\rm GeV}$, the weak neutral-current coupling is suppressed relative to the electromagnetic-current coupling:
\beq
\tan\theta_W\left(\frac{m_X}{m_Z}\right)^2\simeq2.6\times10^{-6}.
\label{eq:zsuppression}
\eeq
The remaining $J_D^\mu J_{Z\mu}$ term is a contact interaction suppressed by $m_Z^{-2}$ and is neglected together with the other $Z$-suppressed contributions, and retain the photon-like interaction for the direct-detection calculation. Note that bare mass terms for $X$ and the dark fermions violate the axial symmetry. In a minimal scalar realization, these masses arise from symmetry breaking, and for $m_{p_D, e_D} > m_X$, this can generate non-perturbative Yukawa couplings. However, as briefly noted in Ref.~\cite{Kaplan:2009de}, there can be other ways of breaking the axial symmetry without coupling it to a perturbative scalar sector. For the purpose of this paper, we treat the broken $U(1)_X$ here as a low-energy realization.

The axial structure is important because it naturally selects transitions between the hyperfine states of the atom.  For an initially neutral atom in its $1s$ singlet ground state, the leading process is
\beq
H_D(S=0)+N\longrightarrow H_D^\ast(S=1)+N .
\label{eq:transition}
\eeq
The detailed spin algebra is collected in
Appendix~\ref{app:spinselection}.

\subsection{Ground-state elastic scattering and the low-velocity population}

A useful consequence of the same interaction is that the much larger
population of low-velocity dark atoms does not generate a leading
conventional elastic recoil spectrum.  For a neutral atom in the hyperfine singlet, the diagonal matrix element vanishes,
\beq
{\cal M}_{0\to0}=0,
\label{eq:elasticzero}
\eeq
whereas the matrix element for a transition from the singlet to the triplet state is nonzero.  This result follows from the spin structure of the axial current rather than from a special choice of constituent masses.

This distinction is worth emphasizing.  A separate cancellation of the first-Born elastic amplitude can occur for degenerate constituent masses in the principal-level excitation mechanism discussed in Ref.~\cite{Kaplan:2009de}.  Our benchmark instead has
$r=10$ and does not make use of that cancellation.

The leading neutral-atom signal is therefore endothermic and must supply the hyperfine excitation energy. The minimum incident speed required to produce a recoil $E_R$ is
\beq
v_{\min}(E_R)=\frac{1}{\sqrt{2m_NE_R}}\left(\frac{m_NE_R}{\mu_{DN}}+\delta_{\rm hf}\right),
\label{eq:inelasticthreshold}
\eeq
where, $\mu_{DN}$ is the reduced mass for the dark atom and target nucleus, given as
\beq
\mu_{DN}\equiv\frac{m_Dm_N}{m_D+m_N},
\eeq

Minimizing this expression with respect to $E_R$ gives the absolute kinematic threshold
\beq
v_{\rm thr}=\sqrt{\frac{2\delta_{\rm hf}}{\mu_{DN}}}.
\eeq
below which the hyperfine transition is forbidden for all recoil energies.

\begin{figure}[t]
\centering
\includegraphics[width=\columnwidth]{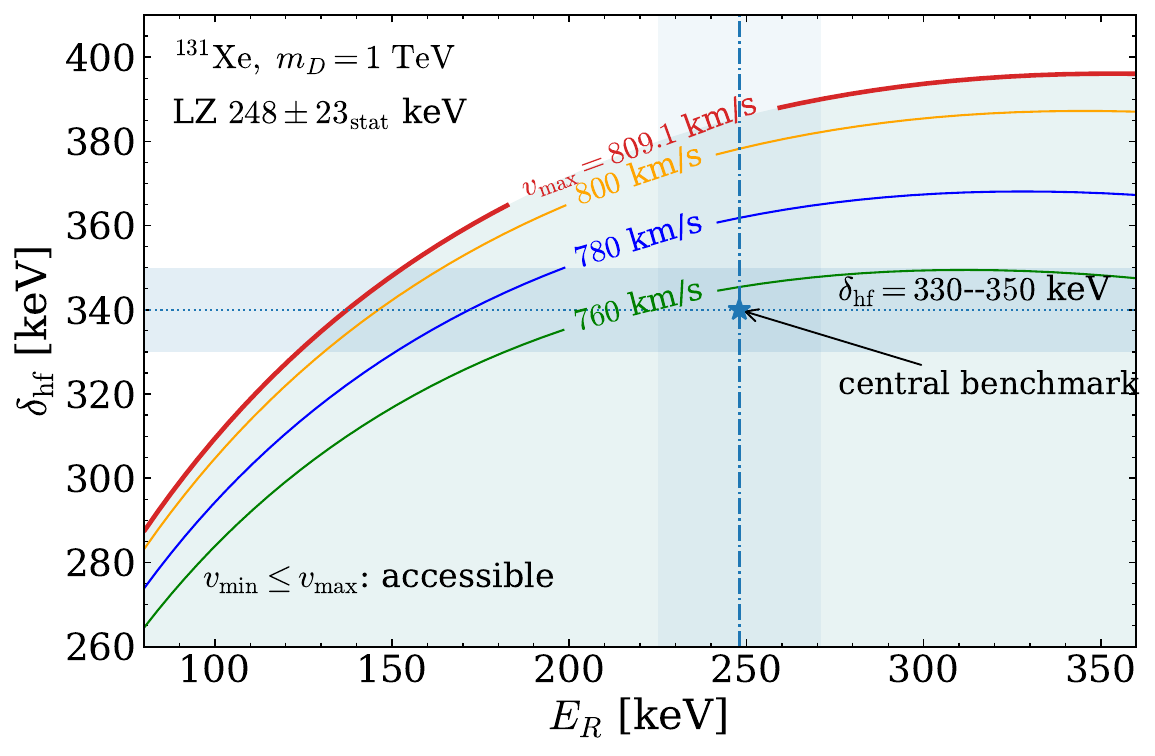}
\caption{Here we demonstrate the kinematic accessibility of endothermic scattering on $^{131}{\rm Xe}$ for $m_D=1~{\rm TeV}$.
The shaded region satisfies $v_{\min}(E_R,\delta_{\rm hf})\leq v_{\max}$, with the red boundary set by the maximum annual SHM speed, $v_{\max}=809.1~{\rm km\,s^{-1}}$. The thin contours indicate $v_{\min}=760$, $780$, and $800~{\rm km\,s^{-1}}$. The horizontal band spans $\delta_{\rm hf}=330$--$350~{\rm keV}$, and the star marks $(E_R,\delta_{\rm hf})=(248,340)~{\rm keV}$.
The vertical line and band show the LZ candidate energy, $248\pm23~({\rm stat})~{\rm keV}$ \cite{LZ:2026axp}, as a visual reference on the true-recoil axis; detector smearing and the systematic energy uncertainty are not included here.}
\label{fig:kinematic_accessibility}
\end{figure}

Equivalently, for an incident speed $v\geq v_{\rm thr}$, the allowed recoil interval is
\beq
E_R^\pm(v)=\frac{\mu_{DN}^2v^2}{2m_N}
\left(1\pm\sqrt{1-\frac{2\delta_{\rm hf}}{\mu_{DN}v^2}}\right)^2,
\label{eq:recoilendpoints}
\eeq
with $E_R^-\leq E_R\leq E_R^+$. In Figure~\ref{fig:kinematic_accessibility} we show the corresponding kinematic region for $^{131}{\rm Xe}$.

The contrast with ordinary elastic scattering is substantial.  If an
unsuppressed elastic interaction were present,
\beq
v_{\min}^{\rm el}(E_R)=\sqrt{\frac{m_NE_R}{2\mu_{DN}^2}}
\simeq21.5~{\rm km\,s^{-1}}\sqrt{\frac{E_R}{\rm keV}}
\label{eq:vminelastic}
\eeq
for $m_D=1~{\rm TeV}$ scattering on Xenon.  Recoils of $5$, $10$, and
$50~{\rm keV}$ would then probe approximately $48$, $68$, and
$152~{\rm km\,s^{-1}}$, respectively, and would sample the bulk of the Galactic halo.

The absence of a leading low-energy recoil population should therefore not be interpreted as an absence of low-velocity dark matter.  The low-velocity neutral component is present, but its leading elastic matrix element vanishes, whereas the hyperfine channel becomes accessible only in the extreme high-velocity tail.  

eq.~\eqref{eq:elasticzero} holds at leading order, but higher-order $X$ exchange, relativistic corrections, additional effective operators, or scattering from a residual ionized component can generate elastic interactions.  In particular, free dark ions do not pay the hyperfine excitation energy and may scatter elastically.  Their local abundance need not coincide with the cosmological residual ionized fraction, and we therefore do not include an ion contribution in the neutral-atom rate below.
This limitation is revisited in Sec.~\ref{sec:cosmology}.

\subsection{Hyperfine transition cross section}

The leading differential cross section for the hyperfine transition is \cite{Kaplan:2009de}
\beq
\frac{d\sigma_{\rm hf}}{dE_R}=\frac{4Z^2\alpha m_N^2}{\mu_{ne_D}^2f_{\rm eff}^4v^2}\,\frac{E_R F_{N}^2(q)F_{\rm el}^2(q)}{
\left(1+2m_NE_R/m_X^2\right)^2}G^2(q),
\label{eq:dsigma}
\eeq
with
\beq
f_{\rm eff}^4=\frac{m_X^4}{2(g_5\epsilon_Yc_W)^2},~~
\mu_{ne_D}=\frac{m_nm_{e_D}}{m_n+m_{e_D}},
\label{eq:feff}
\eeq
and
\beq
F_{\rm el}(q)&=\left(1+\frac{q^2a_D^2}{4}\right)^{-2},\\
G(q)&=1+\frac{\mu_{ne_D}}{m_n}\left[F_{\rm el}^{-1}(q)
-\left(1+F_{\rm el}^{-1}(q)\right)\frac{q^2}{q^2+m_X^2}\right].
\label{eq:formfactors}
\eeq
Here $q=\sqrt{2m_NE_R}$, $Z$ is the nuclear charge, $m_N$ is the
nuclear mass, and $m_n$ is the nucleon mass.  The nuclear response is
described by the Helm form factor $F_{N}(q)$
\cite{Helm:1956zz} for the target nucleus, while $F_{\rm el}(q)$ accounts for the finite size of the dark atom.  

The approximations entering eq.~\eqref{eq:dsigma} neglect terms suppressed
by $m_{e_D}/m_D$ and $m_n/m_N$.  
The factor $E_R\propto q^2$ provides a characteristic momentum weighting of
the hyperfine transition.  It does not by itself imply a spectrum increasing
with recoil energy: the observable shape is controlled jointly by the
mediator propagator, atomic and nuclear form factors, endothermic velocity
integral, and detector response.

For the central benchmark, the LZ candidate at
$E_R=248~{\rm keV}$ \cite{LZ:2026axp} corresponds to
\beq
q\approx 0.246~{\rm GeV}, ~ qa_D\approx 0.045, ~ F_{\rm el}(q)\approx 0.999.
\label{eq:pointlike}
\eeq
The momentum transfer therefore does not resolve the spatial extent of the
dark atom.  Its atomic character enters primarily through the hyperfine
level spacing and the associated transition matrix element.  The xenon
nucleus is not pointlike at these momentum transfers, however, and the
resulting Helm suppression is retained explicitly in
eq.~\eqref{eq:dsigma}.

The massive axial interaction also shifts the hyperfine splitting.  At the leading order\cite{Kaplan:2009de},
\beq
\delta E_{\rm hf}^{(X)}\simeq\frac{g_5^2}{4\pi a_D}\left(1+a_Dm_X\right)^{-2},
\label{eq:portalshift}
\eeq
which, for $m_X=0.20~{\rm GeV}$ and the central atomic benchmark, becomes
\beq
\delta E_{\rm hf}^{(X)}\simeq 0.402\,g_5^2~{\rm GeV}.
\eeq
Requiring this portal-induced contribution to remain below $10\%$ of
$\delta_{\rm hf}=340~{\rm keV}$ gives the working consistency condition
\beq
g_5\lesssim9\times10^{-3}.
\label{eq:g5hyperfine}
\eeq
As shown below, the coupling required by the LZ normalization lies well within this range.
\section{Public LZ recast and portal normalization}
\label{sec:lz}

For natural xenon, the differential recoil rate per unit detector mass is~\cite{Tucker-Smith:2001myb}
\beq
\frac{dR}{dE_R}=\frac{\rho_{D}}{m_D}\sum_i n_{T,i}\int_{v>v_{\min,i}}d^3v\,f_{\rm lab}(\bm v,t)\,v\,\frac{d\sigma_{{\rm hf},i}}{dE_R},
\label{eq:rate}
\eeq
where $i$ labels xenon isotopes, $n_{T,i}$ is their number per unit detector mass, and $\rho_D\equiv f_{D,{\rm loc}}\rho_{\rm DM,loc}$ is the local neutral dark atom density, where in the local neighborhood $\rho_{\rm DM,loc}=0.30~{\rm GeV\,cm^{-3}}$.
Here $f_{D,{\rm loc}}$ is the fraction of the local dark matter mass density in atomic dark atoms, distinct from the cosmological
atomic dark matter fraction $f_D$ introduced in Sec.~\ref{sec:cosmology}.  Since the cross section in eq.~\eqref{eq:dsigma}
is proportional to $v^{-2}$, we define the mean inverse speed as
\beq
\eta(v_{\min},t)\equiv\int_{v>v_{\min}}d^3v\,\frac{f_{\rm lab}(\bm v,t)}{v}.
\label{eq:eta}
\eeq
The same function applies to all isotopes, evaluated at the corresponding $v_{\min,i}$. Within the approximations of eq.~\eqref{eq:dsigma},
the recoil spectrum can be written as
\beq
\begin{aligned}
\frac{dR}{dE_R}&=\frac{\rho_D}{m_D}\sum_i n_{T,i}\,\frac{4Z^2\alpha m_{N_i}^2E_R}{\mu_{ne}^2f_{\rm eff}^4}\\
&\quad\times\frac{F_{N,i}^2(q_i)F_{\rm el}^2(q_i)G_i^2(q_i)}{\left(1+q_i^2/m_X^2\right)^2}\,\eta(v_{\min,i},t),
\end{aligned}
\label{eq:rateexplicit}
\eeq
with $q_i^2=2m_{N_i}E_R$ \cite{Kaplan:2009de}. 

For the spectra and normalizations below, we use the uniform annual average in the standard halo model (SHM),
\beq
\overline{\eta}(v_{\min})\equiv\frac{1}{T}\int_0^{T}dt\,\eta(v_{\min},t).
\label{eq:etaannual}
\eeq
 This is a benchmark averaging prescription; we do not weight the rate by the LZ live-time distribution.

Our recast uses the exposure $\mathcal E=2.84~{\rm tonne\,yr}$ and the efficiency  $\epsilon_{\rm NR}(E_R)$, to the public NR efficiency
curve in D.~S.~Akerib et al.~\cite{LZ:2026axp}.  The reported candidate energy is
\beq
E_{\rm obs}=248\pm23~({\rm stat})\pm23~({\rm sys})~{\rm keV}.
\label{eq:lzenergy}
\eeq
To illustrate energy smearing, we adopt the Gaussian kernel
\beq
{\cal G}(E_{\rm rec};E_R,\sigma_E)=\frac{1}{\sqrt{2\pi}\sigma_E}
\exp\!\left[-\frac{(E_{\rm rec}-E_R)^2}{2\sigma_E^2}\right],
\label{eq:gaussian}
\eeq
with $\sigma_E=23~{\rm keV}$, motivated by the statistical uncertainty in eq.~\eqref{eq:lzenergy}.  The reconstructed spectrum is
\beq
\frac{dN_{\rm sig}}{dE_{\rm rec}}=\mathcal E\int_0^\infty dE_R\,\epsilon_{\rm NR}(E_R)\,
{\cal G}(E_{\rm rec};E_R,\sigma_E)\,\overline{\frac{dR}{dE_R}},
\label{eq:response}
\eeq
where the overbar denotes the annual average. 
The total accepted yield is
\beq
\begin{aligned}
N_{\rm sig}&=\int dE_{\rm rec}\,\frac{dN_{\rm sig}}{dE_{\rm rec}}=\mathcal E\int_0^\infty dE_R\,\epsilon_{\rm NR}(E_R)\,\overline{\frac{dR}{dE_R}}.
\end{aligned}
\label{eq:acceptedyield}
\eeq
The normalized smearing kernel preserves the accepted yield, with no additional cut on $E_{\rm rec}$.
Details of the numerical analysis are given in Appendix~\ref{app:lznumerics}. 


For $m_D=1~{\rm TeV}$ and $\delta_{\rm hf}=340~{\rm keV}$, the maximum annual SHM speed in eq.~\eqref{eq:recoilendpoints} gives
$E_R^-=137.3~{\rm keV}$ for $^{131}{\rm Xe}$. The lower endpoints across natural xenon range from $125.7$ to $159.2~{\rm keV}$,
so the hyperfine recoil spectrum has no support below $125.7~{\rm keV}$ in this halo model. Smearing can populate reconstructed
energies below this edge. 

For the portal benchmark we take
\beq
m_X=0.20~{\rm GeV},\qquad \epsilon_\gamma=4.0\times10^{-4}.
\label{eq:portalbench}
\eeq
At $m_D=1~{\rm TeV}$ and with the atomic parameters specified above, imposing $N_{\rm sig}=1$ gives
\beq
\begin{aligned}
g_5&\simeq1.2\times10^{-3},&&\delta_{\rm hf}=330~{\rm keV},\\
g_5&\simeq1.9\times10^{-3},&&\delta_{\rm hf}=340~{\rm keV},\\
g_5&\simeq3.4\times10^{-3},&&\delta_{\rm hf}=350~{\rm keV}.
\end{aligned}
\label{eq:g5band}
\eeq
 The required coupling rises by nearly a factor of three across this interval, primarily because fewer halo particles can excite the larger splitting. Consequently, the inferred values of $g_5$ are sensitive to the assumed high-speed tail of the halo distribution (See in Sec.~\ref{sec:halo}). These coupling values also respect the consistency condition stated in eq.~\eqref{eq:g5hyperfine}.

\begin{figure}[t]
\centering
\includegraphics[width=\columnwidth]{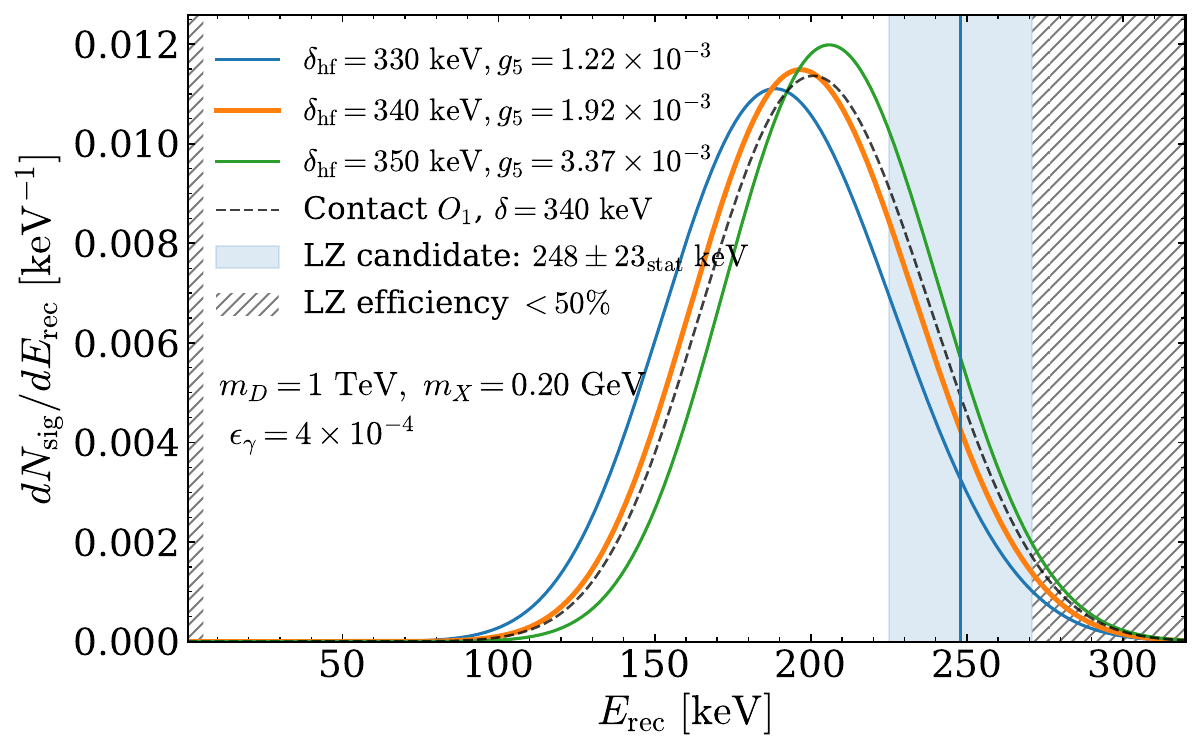}
\caption{Reconstructed natural-xenon event spectra for $m_D=1~{\rm TeV}$, $m_X=0.20~{\rm GeV}$, and $\epsilon_\gamma=4.0\times10^{-4}$. Each solid curve is normalized to $N_{\rm sig}=1$, with $g_5$ given in the legend. The calculation includes the isotope sum, annual SHM average, nuclear and atomic transition factors, selection efficiency, and the Gaussian response in eq.~\eqref{eq:response}. In contrast to elastic or only mildly inelastic scattering, whose spectra are concentrated at lower recoil energies, a sizable endothermic splitting suppresses the low-$E_R$ population and shifts the spectral support to higher recoil energies~\cite{Fan:2026kxx}. This makes the inelastic hyperfine transition particularly well suited to placing appreciable spectral weight near the $248~{\rm keV}$ LZ recoil without simultaneously predicting a much larger population of lower-energy events. The dashed curve shows the contact endothermic $O_1$ reference with $\delta=340~{\rm keV}$ and equal proton and neutron couplings, treated with the same response and independently normalized to one event. The shaded band denotes $248\pm23~{\rm keV}$ using only the candidate's statistical uncertainty~\cite{LZ:2026axp}.}
\label{fig:spectra}
\end{figure}

Figure~\ref{fig:spectra} shows the reconstructed spectra from eq.~\eqref{eq:response}, each normalized to one accepted event, i.e., $N_{\rm sig}=1$.
For these benchmarks, increasing $\delta_{\rm hf}$ shifts the maximum from approximately $189$ to $206~{\rm keV}$. The predicted distribution is broad:
approximately $21\%$ of the accepted yield for $\delta_{\rm hf}=340~{\rm keV}$ lies within the shaded $248\pm23~{\rm keV}$ band; as shaded in blue. The dashed region is not a exclusion region but emphasizes $<50\%$ detector efficiency as reported by  D. S. Akerib et al.~\cite{LZ:2026axp}.   As a reference, we also consider pointlike dark matter with a contact endothermic interaction $O_1=\mathbf 1_{\rm DM}\mathbf 1_N$, equal proton and neutron couplings, $m_{\rm DM}=1~{\rm TeV}$, and $\delta=340~{\rm keV}$ \cite{Tucker-Smith:2001myb}. The finite nuclear size
is retained through the Helm form factor:
\beq
\overline{\frac{dR_{O_1}}{dE_R}}\propto\sum_i n_{T,i}\,m_{N_i}A_i^2F_{N,i}^2(q_i)\,\overline{\eta}(v_{\min,i}),
\label{eq:O1shape}
\eeq
where $A_i$ is the isotope mass number. We apply the same efficiency and smearing and normalize this reference independently to one accepted event. The similar spectra mainly reflect the common endothermic kinematics, nuclear form factor, and detector acceptance. Since $q\simeq0.246~{\rm GeV}$ is comparable to $m_X$ at the candidate recoil energy, the contact limit is not appropriate for the portal. 

For our benchmark parameters, the $X$ mediator is kinematically forbidden from decaying into dark fermions, $\lbrace e_D,p_D\rbrace$,  while it also lies below the dimuon threshold.  Assuming no additional appreciable decay channels, its visible width is therefore dominated by $X\to e^+e^-$.  

At the benchmark mediator mass $m_X=0.20~{\rm GeV}$, we obtain from the
BaBar search~\cite{BaBar:2014zli}
\beq
\epsilon_\gamma<\epsilon_\gamma^{\rm BaBar}\simeq 8.3\times10^{-4}
\qquad (90\%~{\rm C.L.}) .\label{eq:babarreference}
\eeq
The numerical value is obtained by interpolating the digitized BaBar limit provided by DarkCast~\cite{Ilten:2018crw}, as described in
Appendix~\ref{app:lznumerics}. The benchmark kinetic mixing satisfies this constraint.

At the central benchmark, using eq.~\eqref{eq:portalshift}, the coupling required by the one-event normalization yields
\beq
\delta E_{\rm hf}^{(X)}\simeq1.5~{\rm keV},\qquad
\frac{\delta E_{\rm hf}^{(X)}}{\delta_{\rm hf}}\simeq4.4\times10^{-3}.
\label{eq:portalcentral}
\eeq
Thus the $X$ portal-induced correction is below one percent and comfortably satisfies the earlier 10\% consistency criterion. We therefore treat it perturbatively and retain the benchmark $\delta_{\rm hf}$ in the scattering kinematics.

\begin{figure}[t]
\centering
\includegraphics[width=\columnwidth]{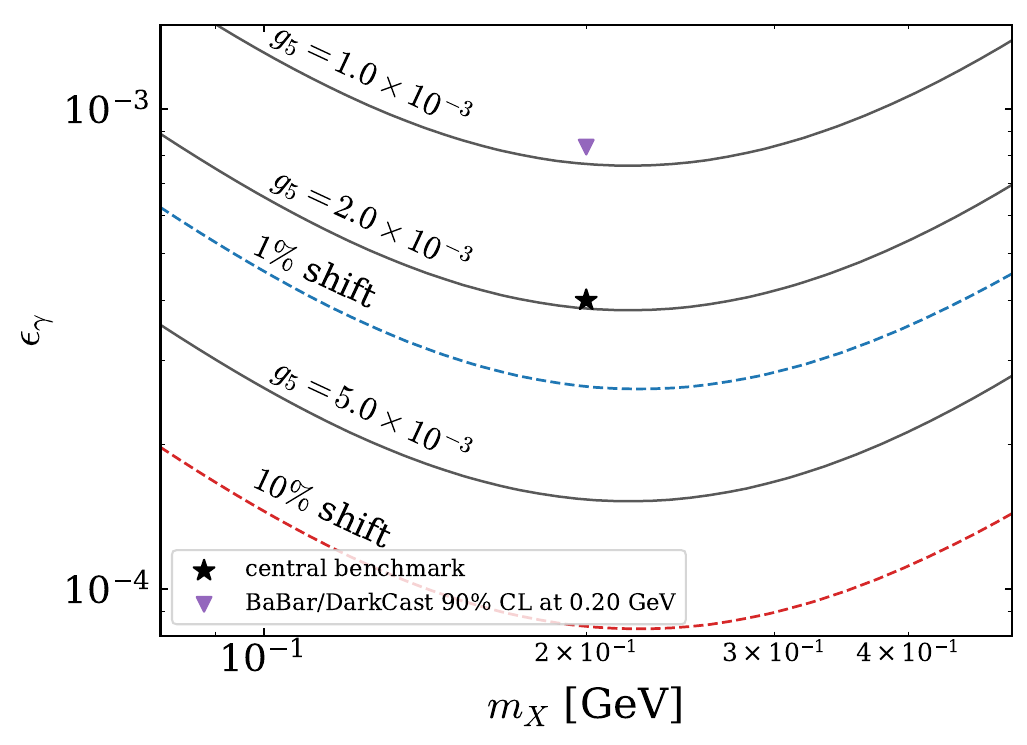}
\caption{In the above we calculate required $g_5$ to produce $N_{\rm sig}= 1$ for  $\delta_{\rm hf}=340~{\rm keV}$ and $m_D=1~{\rm TeV}$; dashed contours give $\delta E_{\rm hf}^{(X)}/\delta_{\rm hf}=1\%$ and $10\%$. The star marks the benchmark in eq.~\eqref{eq:portalbench}, and the triangle marks the interpolated BaBar upper bound at $m_X=0.20~{\rm GeV}$.
At fixed mediator mass, larger $\epsilon_\gamma$ requires smaller $g_5$; the portal correction is smaller above each dashed contour.
The benchmark lies above the $1\%$ contour and below the local BaBar bound. }
\label{fig:portal}
\end{figure}

Figure~\ref{fig:portal} extends the one-event normalization over the $(m_X,\epsilon_\gamma)$ plane. At each point, we determine
$g_5$ from eq.~\eqref{eq:acceptedyield}, as shown in solid contours. The dashed contours delineates $\delta E_{\rm hf}^{(X)}/\delta_{\rm hf}=1\%$ and $10\%$, evaluated with the same coupling.

At $m_X=0.20~{\rm GeV}$, the one-event condition gives $g_5\epsilon_\gamma\simeq7.69\times10^{-7}$.
Requiring the $X$ portal correction below $1\%$ gives $\epsilon_\gamma\gtrsim2.64\times10^{-4}$; the corresponding value for $10\%$
is $8.36\times10^{-5}$. Combining the stricter condition with eq.~\eqref{eq:babarreference} leaves the local interval
\beq
2.6\times10^{-4}\lesssim\epsilon_\gamma\lesssim8.3\times10^{-4},\qquad m_X=0.20~{\rm GeV},
\label{eq:portalinterval}
\eeq
which contains the benchmark in eq.~\eqref{eq:portalbench}. This overlap shows that the coupling required by the recoil
normalization need not appreciably alter the atomic splitting while respecting this BaBar bound. 

\section{Kinematic and astrophysical dependence}
\label{sec:halo}
The endothermic hyperfine transition makes the signal sensitive both to the target nucleus and to the high-velocity structure of the local dark matter distribution. We first examine the target hierarchy, annual modulation, and halo dependence of the signal normalization, before turning to the specific impact of the MW+LMC high-velocity tail.
\subsection{Target hierarchy and halo dependence}
The large hyperfine splitting leads to two characteristic consequences: a strong target hierarchy and enhanced sensitivity to the high-speed halo population. For endothermic scattering, the largest splitting accessible to a nucleus $N$ at speed $v$ is
\beq
\delta_{\rm hf}^{\max}(N)=\frac12\mu_{DN}v^2,
\label{eq:deltamax}
\eeq
from the available center-of-mass kinetic energy \cite{McCabe:2026crm}. For $m_D=1~{\rm TeV}$ and $v_{\max}\simeq809.1~{\rm km\,s^{-1}}$,
\beq
\begin{aligned}
\delta_{\rm hf}^{\max}({}^{40}{\rm Ar})&\simeq131~{\rm keV},&
\delta_{\rm hf}^{\max}({}^{73}{\rm Ge})&\simeq232~{\rm keV},\\
\delta_{\rm hf}^{\max}({}^{131}{\rm Xe})&\simeq396~{\rm keV},&
\delta_{\rm hf}^{\max}({}^{184}{\rm W})&\simeq533~{\rm keV}.
\end{aligned}
\label{eq:targetreach}
\eeq
Thus the $\delta_{\rm hf}=330$--$350~{\rm keV}$ transition is closed on Ar and Ge, lies near the xenon ceiling, and remains open on W. This hierarchy is purely kinematic at fixed available speed and is independent of the portal normalization.

\begin{figure}[t]
\centering
\includegraphics[width=\columnwidth]{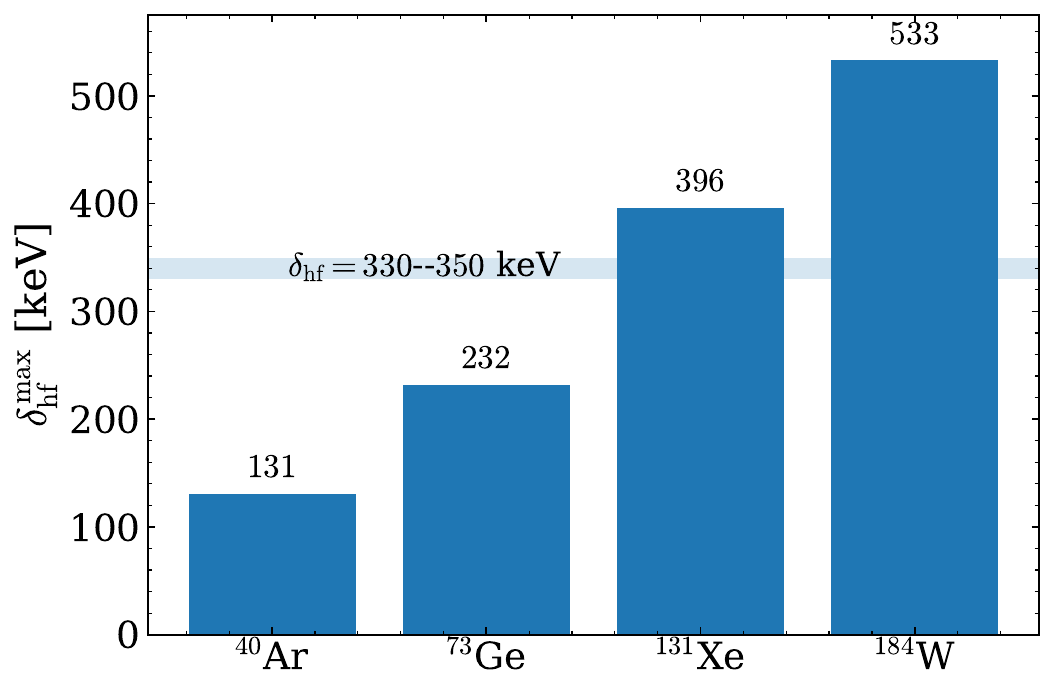}
\caption{Maximum endothermic splitting accessible to representative $^{40}$Ar, $^{73}$Ge, $^{131}$Xe, and $^{184}$W targets at $v_{\max}\simeq809.1~{\rm km\,s^{-1}}$ for $m_D=1~{\rm TeV}$. The shaded band denotes the $\delta_{\rm hf}=330$--$350~{\rm keV}$ interval, which is closed on Ar and Ge, near threshold on Xe, and open on W.}
\label{fig:targets}
\end{figure}

The proximity of the xenon benchmark to its kinematic ceiling also makes the rate sensitive to the extreme high-speed tail. For the SHM reference we use $v_0=238~{\rm km\,s^{-1}}$ and $v_{\rm esc}=544~{\rm km\,s^{-1}}$ \cite{Baxter:2021pqo}, with
\beq
v_E(t)=250.2~{\rm km\,s^{-1}}+14.9~{\rm km\,s^{-1}}\cos[\omega(t-t_0)],
\label{eq:vearth}
\eeq
where $\omega=2\pi/{\rm yr}$ and $t_0$ is near June 2 \cite{DiMauro:2026ldr}. This approximation is sufficient for our halo comparison; a precision modulation analysis near the cutoff would require the full vector Earth velocity and detector live-time distribution.

\begin{figure}[t]
\centering
\includegraphics[width=\columnwidth]{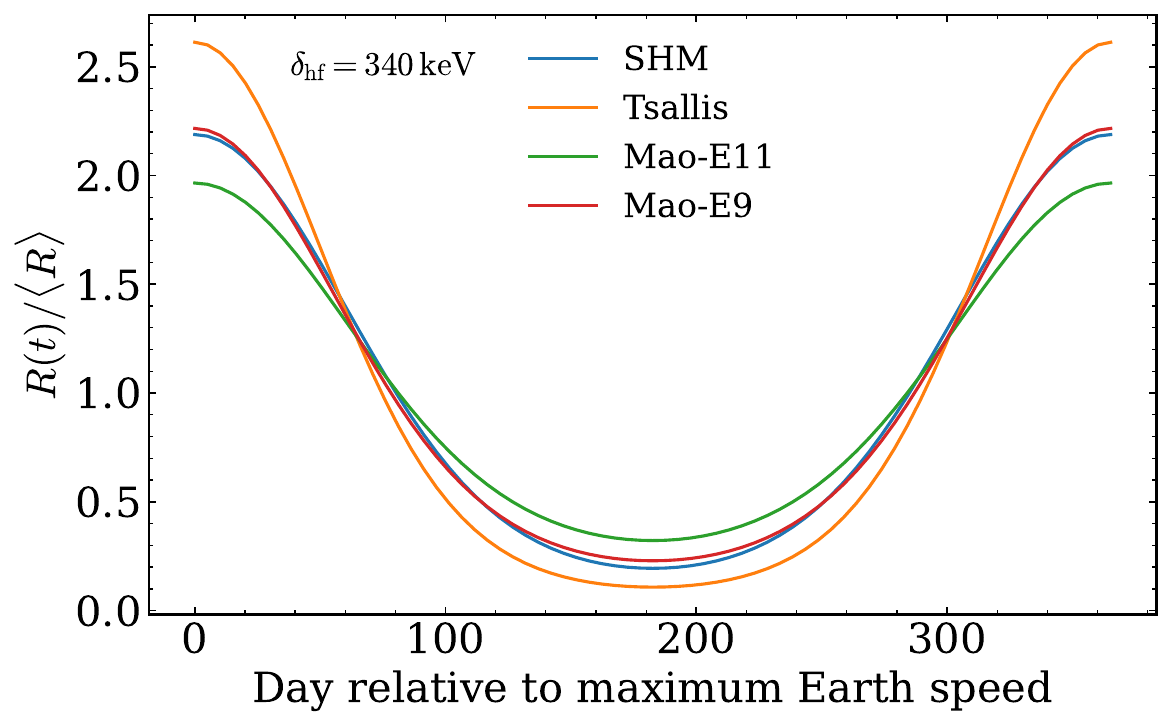}
\caption{Seasonal variation of the accepted recoil rate in xenon for
$\delta_{\rm hf}=340~{\rm keV}$ under the SHM and three representative halo-velocity distributions \cite{Baxter:2021pqo,Ling_2010,Mao:2012mao,Laha:2016iom}.
Day zero is defined as the time of maximum Earth speed, occurring near June~2 in the SHM convention adopted from Ref.~\cite{DiMauro:2026ldr}. Each curve is normalized to its own annual mean. The large modulation arises because the threshold lies near the maximum laboratory-frame speed, making the rate highly sensitive to the high-speed tail.}
\label{fig:halo}
\end{figure}

We also consider a Tsallis-inspired distribution and Mao-type E9 and E11 Milky-Way-like halos as representative variations of the high-speed tail \cite{Ling_2010,Mao:2012mao,Laha:2016iom}. These are stress tests rather than a statistical uncertainty band. Defining
\beq
{\cal A}\equiv\frac{R_{\max}-R_{\min}}{R_{\max}+R_{\min}},
\label{eq:modfrac}
\eeq
the $\delta_{\rm hf}=340~{\rm keV}$ benchmark gives
\beq
\begin{aligned}
{\cal A}_{\rm SHM}&\simeq0.84,& {\cal A}_{\rm Tsallis}&\simeq0.92,\\
{\cal A}_{\rm Mao\text{-}E11}&\simeq0.72,& {\cal A}_{\rm Mao\text{-}E9}&\simeq0.81.
\end{aligned}
\label{eq:modfractions}
\eeq
As can be seen from Figure~\ref{fig:halo}, the precise amplitudes are halo dependent, but the large modulation is common to all four cases because the scattering threshold lies close to the maximum laboratory speed.

The annual mean normalization is even more halo sensitive. Since $N_{\rm sig}\propto(g_5\epsilon_\gamma)^2$,
\beq
\frac{(g_5\epsilon_\gamma)_h}{(g_5\epsilon_\gamma)_{\rm SHM}}
=\left(\frac{\overline R_{\rm SHM}}{\overline R_h}\right)^{1/2}.
\label{eq:halorescale}
\eeq

Relative to the SHM, the required coupling product is enhanced by factors of approximately $4.0$, $1.5$, and $4.1$ for Tsallis, Mao-E11, and Mao-E9, respectively. The SHM value of $g_5$ should therefore not be interpreted as halo independent. Even allowing for these variations, taking $\epsilon_\gamma$ up to the DarkCast-interpolated BaBar reference while requiring $\delta E_{\rm hf}^{(X)}/\delta_{\rm hf}<10\%$ leaves a viable portal window in all four cases; the Tsallis example is the most restrictive. 

\subsection{Impact of the LMC high-velocity tail}

The large recoil energy of the LZ event makes its interpretation particularly sensitive to the high-velocity tail of the local dark matter distribution, where departures from the SHM can be important. The infall of the Large Magellanic Cloud (LMC) provides a well-motivated example, as simulations of MW+LMC analogues find that LMC-associated dark matter can populate the local high-speed tail, while the dynamical response of the MW halo can further boost native halo particles to large velocities~\cite{Smith-Orlik:2023kyl,Reynoso-Cordova:2024xqz}. The resulting MW+LMC distribution therefore extends to larger speeds and enhances the halo integral $\eta_v(v_{\min})$ at the large $v_{\min}$ relevant for highly endothermic scattering~\cite{Bozorgnia:2026emp,Fan:2026kxx}. 

\begin{figure}[t]
    \centering
    \includegraphics[width=1.0\linewidth]{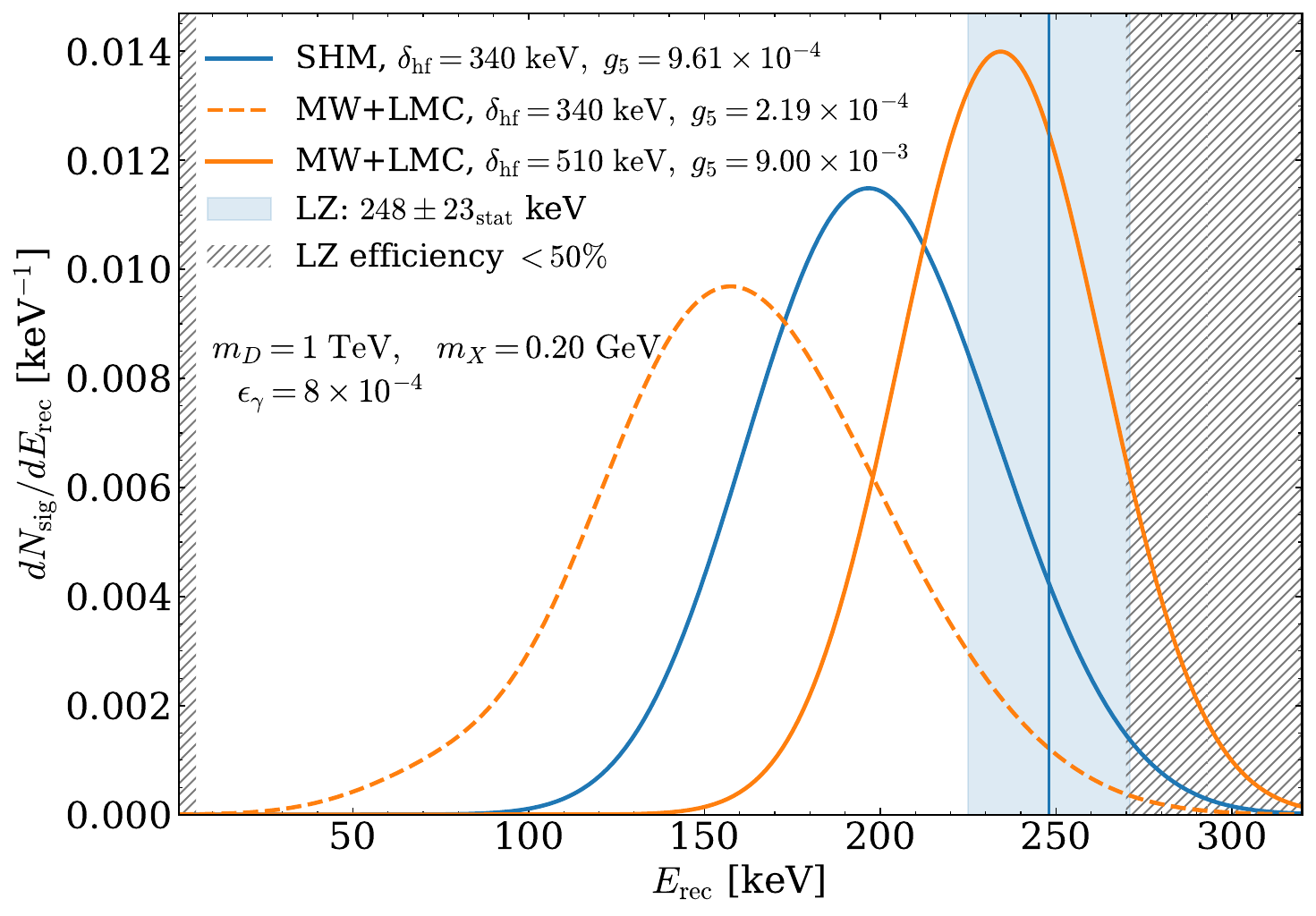}
    \caption{Reconstructed recoil spectra for the SHM and MW+LMC distributions considering the Babar floor, $\epsilon_\gamma = 8\times 10^{-4}$, and normalized to single event. At fixed $\delta_{\rm hf}=340~{\rm keV}$, the shifts MW+LMC the spectrum toward lower recoil energies (dashed orange curve) as compared to the SHM case (solid blue curve), where, the printed $g_5$ values satisfy condition $N_{\rm sig}=1$. However, the MW+LMC high-velocity tail allows for a larger splitting, and for the optimistic case (eq.~\eqref{eq:g5hyperfine}) of $g_5 = 9\times 10^{-3}$, the $N_{\rm sig}=1$ normalization gives $\delta_{\rm hf}\simeq 510~{\rm keV}$, and shifts the spectral support (Solid orange curve) around the LZ $248\pm23~{\rm keV}$ event. Thus, the LMC tail broadens the viable inelastic parameter space rather than merely rescaling the SHM prediction.}\label{fig:LMC}
\end{figure}

 At fixed $\delta_{\rm hf}=340~{\rm keV}$, the relative enhancement is largest at large $v_{\min}$. A larger $v_{\min}$ corresponds to smaller $E_R$ [see eq.~\eqref{eq:inelasticthreshold}]\footnote{For endothermic scattering, $v_{\min}(E_R)$ decreases with $E_R$ below $E_R^\ast=\mu_{DN}\delta_{\rm hf}/m_N$ and increases above it.
For the central ${}^{131}{\rm Xe}$ benchmark, $E_R^\ast\simeq303~{\rm keV}$, so the adopted accepted true-recoil range lies on the decreasing branch.}, so the additional high-velocity population preferentially enhances the low-energy recoil part of the spectrum and shifts the normalized reconstructed distribution toward smaller $E_R$, as shown in Figure~\ref{fig:LMC}. At the same time, however, the extended high-speed tail permits substantially larger endothermic splittings that would otherwise be strongly suppressed or kinematically inaccessible in the SHM [See Figure~\ref{fig:kinematic_accessibility}]. Increasing $\delta_{\rm hf}$ then shifts the reconstructed spectrum back toward higher recoil energies, hence, closer to the observed $248~{\rm keV}$ recoil; particularly, for the benchmark shown in Figure~\ref{fig:LMC}, $\delta_{\rm hf}\simeq510~{\rm keV}$, it peaks around $\approx 234$ keV.

The main effect of the LMC is therefore not simply to enhance the rate, but to extend the kinematic reach of the inelastic transition, i.e., larger hyperfine splittings, resulting in better support for the LZ event.

%

\section{Dark Atomic sector cosmological consistency}
\label{sec:cosmology}

Here we examine whether the atomic dark matter parameters chosen for the LZ interpretation are compatible with the cosmological evolution history. We take the cosmological fraction of the total dark matter density in the atomic dark matter sector to be $f_D\equiv\Omega_{\rm aDM}/\Omega_{\rm DM}=1$ \cite{Bansal:2022qbi,Barron:2026adm}. This does not require the local Galactic population to be entirely neutral. Neutral atoms and residual ions can develop different halo distributions, so the local neutral density entering the direct detection rate is a separate astrophysical quantity \cite{Kaplan:2009de,Cyr-Racine:2012tfp}. We take the total relic abundance as an input and consider the subsequent recombination and decoupling of the atomic sector.

\subsection{Dark radiation and the massive $X$ portal}
\label{sec:Xcosmo}

The recombination and dark acoustic oscillation (DAO) analysis concerns the atomic sector governed by the massless binding photon $A_D$. We define the present-day dark photon-to-CMB temperature ratio as
\begin{equation}
\xi_D\equiv\left.\frac{T_D}{T_\gamma}\right|_{z=0}.
\label{eq:xidef}
\end{equation}
The energy density in the massless dark photon can be expressed through its contribution $\Delta N_D$ to the effective number of relativistic species. For one massless dark photon,
\begin{equation}
\Delta N_D=\frac87\left(\frac{11}{4}\right)^{4/3}\xi_D^4\simeq4.40\,\xi_D^4.
\label{eq:dneff}
\end{equation}
For the range $\Delta N_D=0.01$--$0.15$, this gives $\xi_D\simeq0.218$--$0.430$.

For the portal benchmark $m_X=0.20~{\rm GeV}$ and $\epsilon_\gamma=4.0\times10^{-4}$, $X$ decays to the dark constituents are forbidden, $m_X<2m_{e_D}<2m_{p_D}$, while the di-muon and two-pion thresholds are also closed. The dominant visible mode is therefore $X\to e^+e^-$, with $\Gamma(X\to e^+e^-)\simeq7.8\times10^{-11}~{\rm GeV}$ and $\tau_X\lesssim8.5\times10^{-15}~{\rm s}$. Thus, $X$ does not survive as a thermal relic through big bang nucleosynthesis (BBN) or contribute directly to late-time dark radiation. Its decay and the associated entropy accounting are discussed in Appendix~\ref{app:cosmology}.

The temperature ratio inherited by the dark atomic sector depends on when the energy exchange ceases with the visible sector. A small thermal abundance of $X$ does not by itself suppress scattering through virtual $X$ exchange. We therefore treat $\xi_D$ as a cosmological input and assume that appreciable energy transfer has ceased before dark recombination, with no subsequent reheating of the dark photons. We also assume that additional states associated with $U(1)_X$ breaking make a negligible contribution to the radiation density and that the binding and recombination dynamics are dominated by $A_D$. These assumptions specify the thermal histories to which the following estimates apply. Separate entropy conservation can yield $\xi_D\simeq0.34$--$0.40$, corresponding to $\Delta N_D\simeq0.06$--$0.11$, for the illustrative history described in Appendix~\ref{app:cosmology}; this range is not a prediction of the portal benchmark.

\subsection{Dark recombination and residual ionization}

We adapt the treatment of Cyr-Racine and Sigurdson~\cite{Cyr-Racine:2012tfp} to describe the dark recombination history. We define the ionized fraction of the atomic dark matter plasma as
\begin{equation}
x_D\equiv\frac{n_{e_D}}{n_{p_D}},
\end{equation}
where $n_{e_D}$ is the free dark electron number density and $n_{p_D}$ is the total dark proton number density, including both free dark protons and those bound in dark atoms. At the epoch of dark recombination, $T_D\ll m_{e_D},m_{p_D}$, so both dark fermions are non-relativistic. Within the minimal dark atomic sector, the remaining dark radiation bath therefore consists only of the massless $A_D$. Once its entropy is separately conserved, the dark photon temperature scales as $T_D(z)=\xi_D T_{\gamma,0}(1+z)$, and the corresponding total dark proton number density is
\begin{equation}
n_{p_D}(z)\simeq\frac{\Omega_{\rm aDM}\rho_c}{m_D}(1+z)^3,
\label{eq:ndcosmo}
\end{equation}
where $\rho_c$ is the critical density today.

Using the parameter choice in eq.~\eqref{eq:centralatom}, in particular $m_D=1~{\rm TeV}$ and $B_D\simeq178.5~{\rm MeV}$, together with $\Omega_{\rm aDM}h^2\simeq0.12$, the Saha relation in Appendix~\ref{app:cosmology} gives
\begin{equation}
T_D^{\rm rec}\simeq4.7\text{--}5.0~{\rm MeV},\qquad B_D/T_D^{\rm rec}\simeq36\text{--}38,
\label{eq:trec}
\end{equation}
where $x_{D,\rm Saha}=1/2$ is used only as a marker for the onset of recombination. Over the same range of $\xi_D$, this corresponds to
\begin{equation}
z_{\rm rec}\simeq(5\text{--}10)\times10^{10}.
\label{eq:zrec}
\end{equation}
The Saha criterion marks the onset of recombination rather than its freeze-out; the surviving ionized fraction is set by the subsequent departure from ionization equilibrium.

For a late-time analytic estimate of the residual ionization fraction, we use~\cite{Cyr-Racine:2012tfp}
\begin{equation}
\begin{aligned}
\bar x_D\sim{}&1.8\times10^{-16}\frac{\xi_D}{\alpha_D^6}\left(\frac{\Omega_{\rm aDM}h^2}{0.12}\right)^{-1}\\
&\times\left(\frac{m_D}{\rm GeV}\right)\left(\frac{B_D}{\rm keV}\right).
\end{aligned}
\label{eq:xres}
\end{equation}
Here $\bar x_D$ denotes the asymptotic ionized fraction after dark recombination freezes out. With the central parameter choices in eq.~\eqref{eq:centralatom}, and the range of $\xi_D$ above, eq.~\eqref{eq:xres} gives
\beq
\bar x_D\simeq0.09\text{--}0.17.
\label{eq:xresrange}
\eeq
This analytic estimate suggests a predominantly neutral cosmological population. It is not a precision recombination calculation: Ref.~\cite{Cyr-Racine:2012tfp} notes that the estimate can differ from the numerical result by as much as an order of magnitude. More recently, Barron et al.~\cite{Barron:2026adm} found that the recombination coefficients and atomic transition rates relevant for cosmological dark recombination agree with first-principles calculations at approximately the $\mathcal O(10\%)$ level over the mass-ratio and coupling regime containing our benchmark. This validates the rate coefficients rather than the accuracy of the analytic freeze-out estimate. We therefore use eq.~\eqref{eq:xresrange} to estimate the surviving charged fraction to the order of magnitude.

\subsection{Dark acoustic decoupling}

Residual free dark electrons and protons exchange momentum with the dark photon bath through Compton scattering. In the low-energy, nonrelativistic limit, the physical drag rate for the combined massive component is~\cite{Cyr-Racine:2012tfp}
\beq
\Gamma_{\rm drag}^{({\rm C})}(a)\simeq\frac{4\rho_{\gamma_D}(a)}{3m_D}\,x_D(a)\sigma_{T,D},
\label{eq:drag}
\eeq
where $\rho_{\gamma_D}=\pi^2T_D^4/15$ and $\sigma_{T,D}\simeq1.7\times10^{-33}~{\rm cm^2}$ is the dark Thomson cross section. The opacity, the distinction between photon decoupling and matter drag, and the approximations entering the scale estimate are given in Appendix~\ref{app:cosmology}.

Approximating $x_D(a_{\rm drag})\approx\bar x_D$, we estimate the dark photon temperature at kinetic decoupling by solving $\Gamma_{\rm drag}^{({\rm C})}(a_{\rm drag})=H(a_{\rm drag})$, obtaining
\beq
T_D^{\rm drag}\simeq0.045\text{--}0.13~{\rm MeV}.
\label{eq:Tdrag}
\eeq
This is well below $T_D^{\rm rec}\simeq5~{\rm MeV}$ and is consistent with using a residual-ionization estimate at the drag epoch. Evaluating the Rayleigh and photoionization opacities of Ref.~\cite{Cyr-Racine:2012tfp} at these temperatures shows that both contributions are negligible compared with Compton scattering for our benchmark, as detailed in Appendix~\ref{app:cosmology}. Using the effective sound-horizon integral in eq.~\eqref{eq:rdao} as a characteristic comoving scale gives
\begin{equation}
r_{\rm DAO}^{({\rm C})}\simeq1.1\times10^{-4}\text{--}6.0\times10^{-4}~{\rm Mpc}
\label{eq:rdaorange}
\end{equation}
over $\Delta N_D=0.01$--$0.15$. As a conservative stress test of the Compton-drag estimate, setting $x_D=1$ throughout the calculation gives
\begin{equation}
r_{\rm DAO}^{({\rm C})}<1.5\times10^{-3}~{\rm Mpc}.
\label{eq:rdaoconservative}
\end{equation}
This fully ionized case is not a physical post-recombination history, but provides an upper envelope within the same drag prescription.

Using Planck and ACT CMB data together with BAO and Pantheon+, Ref.~\cite{Barron:2026adm} finds the one-dimensional marginalized 95\% limits
\begin{equation}
r_{\rm DAO}<2.4~{\rm Mpc},\qquad \Delta N_D<0.15,
\label{eq:barronlimits}
\end{equation}
for $f_D=1$ and $m_{p_D}=1~{\rm TeV}$, the nearest tabulated benchmark to our model. Even the fully ionized stress test lies more than three orders of magnitude below this reference DAO limit.  A full CLASS-aDM treatment~\cite{Bansal:2022qbi} would determine the transfer function and the precise decoupling history; the scale comparison here is intended as a consistency estimate for the assumed thermal history.

For the LZ interpretation, the cosmological abundance must be distinguished from the local neutral density used in the recoil rate. The small neutral self-scattering estimates in Appendix~\ref{app:cosmology} support a weakly self-interacting neutral component. Residual ions can have a different Galactic distribution and can scatter through the $X$ portal without the neutral-atom hyperfine threshold. Their recoil contribution depends on their local abundance and portal couplings and is not fixed by eq.~\eqref{eq:xresrange} alone.
\section{Conclusions}
\label{sec:conclusions}

Hydrogen-like atomic dark matter provides a microscopic realization of inelastic scattering through its hyperfine spectrum \cite{Kaplan:2009de,Cline:2021itd}. The constituent masses and binding interaction determine the atomic binding energy, size, and hyperfine structure. For the axial interaction considered here, the leading nonrelativistic elastic matrix element of the singlet ground state vanishes, while hyperfine excitation remains allowed. The low-velocity neutral population therefore cannot produce the usual leading elastic recoil spectrum, and the excitation threshold suppresses low-energy nuclear recoils. 

We have examined this mechanism as an interpretation of the high-energy LZ nuclear-recoil candidate. For a dark atom mass of $1~{\rm TeV}$ and a constituent mass ratio, $r\sim 10$, an assumed hyperfine splitting of $330$--$350~{\rm keV}$ selects the high-speed tail of the SHM. Approximately $21\%$ of the central benchmark spectrum lies in the $225$--$271~{\rm keV}$ candidate band specific to the SHM halo model. This fraction describes spectral placement and carries no claim of statistical significance or preference for the benchmark. The interpretation is particularly sensitive to the high-speed velocity distribution. Our comparison of the SHM with the Tsallis and Mao distributions shows substantial changes in the required normalization and annual modulation, consistent with earlier studies of halo uncertainties in direct detection \cite{Ling_2010,Mao:2012mao,Laha:2016iom}. These effects must be included when relating a high-energy recoil to microscopic dark-sector parameters.

The MW+LMC distribution provides a further possibility. Dark matter associated with the LMC and the dynamical response of the MW halo can extend the local high-speed population, increasing the accessible endothermic splitting \cite{Smith-Orlik:2023kyl,Reynoso-Cordova:2024xqz}. In the example considered here, a splitting of $\delta_{\rm hf}\simeq 510~{\rm keV}$ places the reconstructed spectral maximum close to the $248~{\rm keV}$ candidate. The comparison with SHM under single event normalization demonstrates additional kinematic reach and a possible recoil scale.

We find that the excitation threshold also produces a pronounced dependence on the detector target. For the adopted SHM and splitting range, the hyperfine channel is kinematically closed on Ar and Ge, accessible near threshold on Xe, and open on W. Xe and W therefore offer complementary tests of this benchmark through searches extending to high recoil energies. The target hierarchy depends on the dark atom mass, splitting, and available halo speeds, while the observable rates also depend on the interaction, nuclear response, exposure, and detector acceptance.

In the context of limits from solar-neutrino observation, the solar gravitational acceleration can make hyperfine excitation accessible on heavy nuclei such as Ca, Fe, and Ni.
For $\delta_{\rm hf}=340~{\rm keV}$ and a local dark matter speed of approximately $1400~{\rm km\,s^{-1}}$ inside the Sun, hyperfine excitation requires a dark matter nucleus reduced mass $\mu_{DN}\gtrsim31~{\rm GeV}$, while capture additionally requires sufficient energy loss.
We stress that solar-neutrino limits from dark matter annihilations \cite{DiMauroShaikh:2026solar, Bose:2026ndd} do not apply to our assumed asymmetric atomic sector.
We leave a dedicated calculation of capture and subsequent atomic-state evolution for future work. Our estimates of dark recombination, radiation drag, and neutral self-scattering provide cosmological checks under the stated thermal history assumptions.
A complete assessment must also establish the local neutral abundance and consistently include the axial portal's contribution to the atomic spectrum. 
We think, the connection between atomic excitation, high-energy recoils, target dependence, and annual modulation, is a good motivation for a more careful analysis for testing the atomic interpretation with future data.

\begin{acknowledgments}
PC, BD, and SM thank Nassim Bozorgnia, James Unwin, and Jie Hu for valuable discussions. PC particularly thanks Nassim Bozorgnia for financial support and assistance with the MW+LMC halo data. This research was undertaken thanks in part to funding from the Natural Sciences and Engineering Research Council of Canada through the Arthur B.~McDonald Canadian Astroparticle Physics Research Institute. OpenAI tools aided in the analysis.
\end{acknowledgments}

\appendix

\section{Spin structure of the axial atomic transition}
\label{app:spinselection}

We describe here the spin algebra that specifies underlying
eq.~\eqref{eq:elasticzero}.  The $1s$ hyperfine singlet is
\beq
|0,0\rangle=\frac{|\uparrow_{e_D}\downarrow_{p_D}\rangle-|\downarrow_{e_D}\uparrow_{p_D}\rangle}{\sqrt2},
\label{eq:singletstate}
\eeq
The leading nonrelativistic Hamiltonian is
linear in the individual constituent spins and can schematically be written
as \cite{Kaplan:2009de}
\beq
{\cal H}_{X}^{\rm NR}=\bm C_e\cdot\bm S_{e_D}+\bm C_p\cdot\bm S_{p_D},
\label{eq:NRspinHamiltonian}
\eeq
The singlet satisfies
\beq
\bm S_{e_D}|0,0\rangle=\frac{|\uparrow_{e_D}\downarrow_{p_D}\rangle+|\downarrow_{e_D}\uparrow_{p_D}\rangle}{\sqrt2}= | 1, 0 \rangle~,
\label{eq:diagonalspinzero}
\eeq
Hence,
\begin{equation}
    \langle 0, 0| \bm S_{e_D} |0, 0 \rangle =  \langle 0, 0 | 1, 0 \rangle = 0~.
\end{equation}
It follows immediately that
\beq
\langle0,0|{\cal H}_{X}^{\rm NR}|0,0\rangle=0,
\eeq
Hence
\beq
{\cal M}_{0\to0}=0,\qquad\sigma_{\rm el}=0,
\eeq
while the transition matrix element from the singlet to the triplet state remains non-zero.

The vanishing diagonal amplitude discussed here should not be confused with
the separate first-Born cancellation considered in
Ref.~\cite{Kaplan:2009de} for transitions between principal atomic levels.
There, taking $r=1$ can cancel the elastic form factor, whereas
the present mechanism relies on the hyperfine spin structure and remains
applicable for the unequal constituent masses used in our benchmark.

\section{Numerical implementation of the LZ signal}
\label{app:lznumerics}

This appendix summarizes the numerical implementation used to obtain the kinematic-reach, recoil-spectrum, portal-normalization, and halo-dependence results presented in the main text. We specify the halo model, xenon isotope
treatment, nuclear form factor, detector efficiency, and numerical normalization procedure. The underlying differential rate is given in eqns.~\eqref{eq:dsigma}--\eqref{eq:acceptedyield}. 

\subsection{LZ signal reconstruction}

For the Standard Halo Model (SHM) we consider~\cite{DiMauro:2026ldr}
\beq
v_0&=238~{\rm km\,s^{-1}},~~
v_{\rm esc}=544~{\rm km\,s^{-1}},\\
v_E(t)&=250.2+14.9\cos\phi~{\rm km\,s^{-1}},
\label{eq:app_shm}
\eeq
where $\phi$ is the orbital phase. The maximum laboratory-frame speed is
therefore
\beq
v_{\max}=v_{\rm esc}+v_E^{\max}
=809.1~{\rm km\,s^{-1}}.
\eeq

For numerical evaluation of the SHM mean inverse speed defined in
eq.~\eqref{eq:eta}, we use its standard analytic form \cite{Lee:2013xxa}.  Defining
\beq
x=\frac{v_{\min}}{v_0},\qquad
y=\frac{v_E}{v_0},\qquad
z=\frac{v_{\rm esc}}{v_0},
\eeq
one has
\beq
\eta'=
\begin{cases}
\begin{aligned}
&{\rm erf}(x+y)-{\rm erf}(x-y)\\[-2pt]
&-\dfrac{4y}{\sqrt{\pi}}e^{-z^2},
\end{aligned}
& x<z-y,\\[8pt]
\begin{aligned}
&{\rm erf}(z)-{\rm erf}(x-y)\\[-2pt]
&-\dfrac{2(z-x+y)}{\sqrt{\pi}e^{z^2}},
\end{aligned}
& z-y\leq x<z+y,\\[6pt]
0, & x\geq z+y ,
\end{cases}
\label{eq:app_eta}
\eeq
with
\beq
\eta' \equiv 2N_{\rm esc}v_E\eta(v_{\min},t),~~N_{\rm esc}={\rm erf}(z)-\frac{2z}{\sqrt{\pi}}e^{-z^2}.
\eeq

For the time-averaged recoil spectra and portal normalization we use the annual-average mean inverse speed defined in
eq.~\eqref{eq:etaannual}, evaluated numerically using 72 uniformly spaced orbital phases.

The finite nuclear size is included through the Helm form factor
\beq
F_N(q)=3\frac{j_1(qR_1)}{qR_1}
\exp\!\left[-\frac{(qs)^2}{2}\right],
\qquad
R_1=\sqrt{R^2-5s^2},
\label{eq:app_helm}
\eeq
where $R_1$ is the effective nuclear radius and $s$ parametrizes the
nuclear surface thickness.  We take
\beq
R=1.2A^{1/3}~{\rm fm},
\qquad
s=0.9~{\rm fm}.
\eeq

The LZ high-energy nuclear-recoil selection efficiency $\epsilon_{\rm NR}(E_R)$ is approximated from a digitization of the public efficiency curve of Ref.~\cite{LZ:2026axp} and applying piecewise-linear interpolation in true
recoil energy.  The interpolation is set to zero outside the digitized range, which extends to approximately $E_R=300~{\rm keV}$.
For a given halo model $h$, the accepted rate is evaluated as
\beq
R_h(t)=\int dE_R\,\epsilon_{\rm NR}(E_R)\frac{dR_h(E_R,t)}{dE_R},
\label{eq:app_Rtime}
\eeq
with all particle-physics parameters and detector inputs held fixed.

The reconstructed spectrum is obtained from eq.~\eqref{eq:response}. We use the Gaussian response in eq.~\eqref{eq:gaussian} with $\sigma_E=23~{\rm keV}$, corresponding to the quoted statistical uncertainty of the high-energy candidate.

\subsection{One-event portal normalization}

The total accepted signal yield is evaluated using
eq.~\eqref{eq:acceptedyield} with the exposure
\beq
\mathcal E=2.84\times10^3\times365.25~{\rm kg\,day}.
\eeq
For the benchmark recoil calculation we take
$f_{D,{\rm loc}}=1$ and
$\rho_{\rm DM,loc}=0.30~{\rm GeV\,cm^{-3}}$ as in
Sec.~\ref{sec:lz}.  

At fixed $(m_X,\delta_{\rm hf})$, eq.~\eqref{eq:dsigma} implies
\beq
N_{\rm sig}\propto(g_5\epsilon_Y)^2.
\label{eq:app_scaling}
\eeq
 After fixing the coupling through the one-event normalization, the accepted true-recoil spectrum is convolved with the detector response according to eq.~\eqref{eq:response}. 
For the BaBar constraint, we use the digitized 90\% C.L. upper-limit curve distributed with
DarkCast~\cite{Ilten:2018crw} as \texttt{BaBar\_Lees2014xha.lmt}, corresponding to the BaBar search of Ref.~\cite{BaBar:2014zli}.  At the benchmark mediator mass
$m_X=0.20~{\rm GeV}$, this procedure gives the value quoted in eq.~\eqref{eq:babarreference}.

\subsection{Dark matter halo models}
Here we summarize the dark matter halo models used in our work.

For the SHM,
\beq
P_{\rm SHM}(u)\propto u^2\exp\!\left[-\left(\frac{u}{v_0}\right)^2\right]\Theta(v_{\rm esc}-u),
\label{eq:app_shm_pdf}
\eeq
with the parameters of eq.~\eqref{eq:app_shm}.

For the Tsallis-inspired profile we adopt the fit of Ref.~\cite{Ling_2010},
\beq
P_{\rm Tsallis}(u)\propto u^2\left[1-(1-q)\left(\frac{u}{v_0}\right)^2\right]^{q/(1-q)}\Theta(v_{\rm esc}-u),
\label{eq:app_tsallis}
\eeq
with
\beq
q=0.773,~~v_0=267.2~{\rm km\,s^{-1}},~~v_{\rm esc}=560.8~{\rm km\,s^{-1}}.
\eeq
The distribution is set to zero when the quantity in square brackets becomes non-positive.

For the Mao-type profiles we use the empirical form of Ref.~\cite{Mao:2012mao},
\beq
P_{\rm Mao}(u)\propto u^2\exp\!\left(-\frac{u}{v_0}\right)
\left[1-\left(\frac{u}{v_{\rm esc}}\right)^2\right]^p
\Theta(v_{\rm esc}-u).
\label{eq:app_mao}
\eeq
For the two representative MW-like halos E9 and E11 we use the Mao-profile fits reported in ~\cite{Laha:2016iom},
\beq
{\rm Mao\mbox{-}E11}:~&v_0=250.06~{\rm km/s},~p=3.14,~v_{\rm esc}=600~{\rm km/s},\\
{\rm Mao\mbox{-}E9}:~&v_0=393.63~{\rm km/s},~p=4.82,~v_{\rm esc}=600~{\rm km/s}.
\label{eq:app_mao_params}
\eeq

\section{The dark atomic-sector cosmology}
\label{app:cosmology}

\subsection{Dark recombination history}

While ionization equilibrium holds, the Saha relation is
\begin{equation}
\frac{x_{D,\rm Saha}^2}{1-x_{D,\rm Saha}}=\frac{1}{n_{p_D}}\left(\frac{m_{e_D}m_{p_D}T_D}{2\pi m_D}\right)^{3/2}e^{-B_D/T_D}.
\label{eq:saha}
\end{equation}
Together with eq.~\eqref{eq:ndcosmo} and $T_D=\xi_D T_{\gamma,0}(1+z)$, this gives eqns.~\eqref{eq:trec} and \eqref{eq:zrec}. An independent analytic criterion for efficient bound-state formation is~\cite{Cyr-Racine:2012tfp}
\begin{equation}
\frac{\alpha_D^6}{\xi_D}\left(\frac{\Omega_{\rm aDM}h^2}{0.12}\right)\left(\frac{m_D}{\rm GeV}\right)^{-1}\left(\frac{B_D}{\rm keV}\right)^{-1}\gtrsim1.5\times10^{-16}.
\label{eq:atomformation}
\end{equation}

Separate conservation of dark photon entropy implies $T_D\propto a^{-1}$, but the ratio $T_D/T_\gamma$ changes during visible-sector entropy release. With the visible entropy degrees of freedom defined relative to the photon temperature,
\begin{equation}
\frac{T_D}{T_\gamma}=\xi_D\left[\frac{g_{*s}^{\rm vis}(T_\gamma)}{g_{*s,0}^{\rm vis}}\right]^{1/3},\qquad g_{*s,0}^{\rm vis}\simeq3.91.
\label{eq:xiT}
\end{equation}

\subsection{Opacity and the characteristic drag scale}

Following Ref.~\cite{Cyr-Racine:2012tfp}, the conformal Compton opacity is
\begin{equation}
\begin{aligned}
\tau_{\rm C}^{-1}&=a n_{p_D}x_D(a)\sigma_{T,D}\left[1+\left(\frac{m_{e_D}}{m_{p_D}}\right)^2\right],\\
\sigma_{T,D}&=\frac{8\pi\alpha_D^2}{3m_{e_D}^2}\simeq1.7\times10^{-33}~{\rm cm^2}.
\end{aligned}
\label{eq:comptonopacity}
\end{equation}
Here $\sigma_{T,D}$ is the dark Thomson cross section, appropriate to the low-energy limit of Compton scattering. The collision term in the massive-component Euler equation is proportional to $(4\rho_{\gamma_D}/3\rho_{\rm aDM})\tau_{\rm C}^{-1}$. 

The total opacity also receives contributions from Rayleigh scattering~\cite{Cyr-Racine:2012tfp},
\begin{equation}
\tau_R^{-1}\simeq32\pi^4 a n_{p_D}(1-x_D)\sigma_{T,D}\left(\frac{T_D}{B_D}\right)^4,
\label{eq:rayleighopacity}
\end{equation}
valid for $T_D\ll B_D$. At the estimated drag epoch,
\begin{equation}
\frac{\tau_R^{-1}}{\tau_C^{-1}}\simeq\frac{32\pi^4(1-x_D)}{x_D[1+(m_{e_D}/m_{p_D})^2]}\left(\frac{T_D}{B_D}\right)^4\lesssim10^{-8}.
\label{eq:rayleighratio}
\end{equation}
 Compton scattering therefore controls the dark photon momentum transfer at the estimated drag epoch to excellent accuracy.

 We use the corresponding integral extended to the drag epoch to define the effective scale
\begin{equation}
r_{\rm DAO}^{({\rm C})}\equiv\int_0^{a_{\rm drag}}\frac{c\,da}{\sqrt{3(1+\mathcal R_D(a))}\,a^2H(a)},
\label{eq:rdao}
\end{equation}
where, $\mathcal R_D(a)\equiv 3\rho_{\rm aDM}/4\rho_{\gamma_D}$.

This definition provides a characteristic scale for the analytic comparison, rather than a calculation of the oscillation pattern or damping in the matter transfer function.

\begin{figure}[t]
\centering\includegraphics[width=\columnwidth]{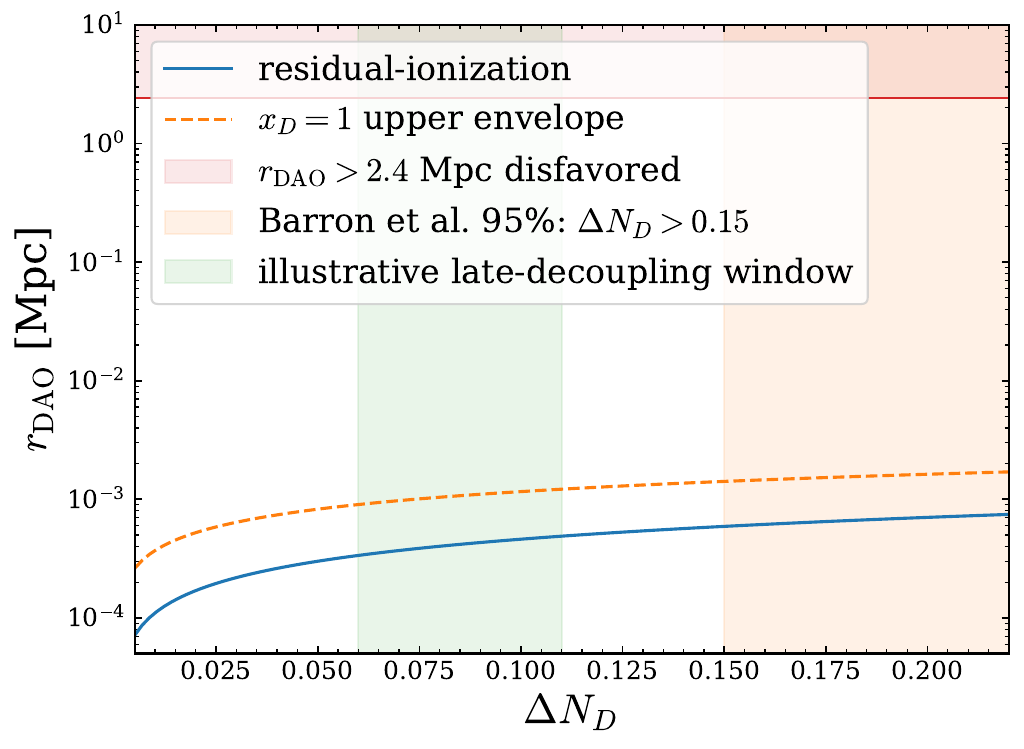}
\caption{ The solid curve uses the residual-ionization estimate in eq.~\eqref{eq:xres}; the dashed curve forces $x_D=1$ to provide an upper envelope. }
\label{fig:cosmology}
\end{figure}

\subsection{Neutral self-scattering}

The compact size of the benchmark atom gives a small geometric self-scattering estimate,
\begin{equation}
\frac{\pi a_D^2}{m_D}\simeq2.3\times10^{-8}~{\rm cm^2\,g^{-1}},
\label{eq:geomself}
\end{equation}
while $\sigma_{HH}\sim4\pi(\kappa a_D)^2$ with $\kappa=3$--$10$ \cite{Kaplan:2009de} gives
\begin{equation}
\frac{\sigma_{HH}}{m_D}\sim8\times10^{-7}\text{--}9\times10^{-6}~{\rm cm^2\,g^{-1}}.
\label{eq:selfanalytic}
\end{equation}
These estimates suggest that neutral self-scattering is comfortably below halo and cluster constraints, although a detailed cross section can depend on velocity and resonances \cite{Cline:2014scatter}. Here we have study the neutral atomic component and do not constrain Coulomb interactions among residual ions or additional interactions mediated by $X$.

\subsection{Portal decay and illustrative entropy histories}

For the benchmark $m_X=0.20~{\rm GeV}$ and $\epsilon_\gamma=4.0\times10^{-4}$, the visible decay width is
\beq
\Gamma(X\to e^+e^-)&=\frac{\alpha\epsilon_\gamma^2m_X}{3}\left(1+\frac{2m_e^2}{m_X^2}\right)\sqrt{1-\frac{4m_e^2}{m_X^2}}\\
&\simeq7.8\times10^{-11}~{\rm GeV},
\label{eq:Xdecay}
\eeq
and hence
\begin{equation}
\tau_X\lesssim8.5\times10^{-15}~{\rm s}.
\label{eq:Xlifetime}
\end{equation}
Rapid decays and inverse decays keep $X$ near its visible-sector equilibrium abundance as it becomes nonrelativistic. This abundance becomes Boltzmann suppressed below $m_X$ and is negligible by the onset of BBN. The associated visible entropy release occurs before neutrino decoupling, so the thermal $X$ population itself gives no late contribution to $N_{\rm eff}$. 

If the visible and dark sectors share a common temperature at decoupling, separate conservation of their comoving entropies subsequently gives the present dark-to-photon temperature ratio
\beq
\xi_D=\left(\frac{g_{*s,0}^{\rm vis}}{g_{*s,{\rm dec}}^{\rm vis}}\right)^{1/3}\left(\frac{g_{*s,{\rm dec}}^D}{g_{*s,0}^D}\right)^{1/3}.
\label{eq:xientropy}
\eeq

As an illustrative history, for the case of dark sector decoupling after the heavy dark fermions have become nonrelativistic and their entropy release is complete. The minimal dark radiation bath then contains only $A_D$, giving $g_{*s,{\rm dec}}^D=g_{*s,0}^D=2$. Using $g_{*s,0}^{\rm vis}\simeq3.91$ and $g_{*s,{\rm dec}}^{\rm vis}\simeq60$--$100$ gives
\begin{equation}
\xi_D\simeq0.34\text{--}0.40,\qquad \Delta N_D\simeq0.06\text{--}0.11,
\label{eq:xilate}
\end{equation}
consistent with the reference radiation bound in eq.~\eqref{eq:barronlimits}. 

For comparison, if the sectors separate while both dark Dirac constituents are relativistic and their subsequent entropy is deposited in the $A_D$ bath, then, neglecting the $U(1)_X$-breaking sector,
\begin{equation}
\frac{g_{*s}^D(T_{\rm dec})}{g_{*s}^D(T_0)}=\frac{2+\frac78(4+4)}{2}=\frac92.
\label{eq:gdofearly}
\end{equation}
For decoupling above the electroweak scale, taking $g_{*s,{\rm dec}}^{\rm vis}=106.75$ gives
\begin{equation}
\xi_D\simeq0.55,\qquad \Delta N_D\simeq0.40,
\label{eq:xiearly}
\end{equation}
which exceeds the reference bound in eq.~\eqref{eq:barronlimits}. The relevant cosmological question is therefore not the late abundance of $X$, which is negligible in the assumed thermal history, but when energy exchange between the visible and dark sectors ceases and where the subsequent entropy is deposited.

\bibliography{refs_DD_LZ_atomic-DM}
\end{document}